\def\cm2{cm$^{-2}$}

\def\c2{C~{\sc ii}}
\def\c4{C~{\sc iv}}
\def\fe2{Fe~{\sc ii}}
\def\fe3{Fe~{\sc iii}}
\def\mg1{Mg~{\sc i}}
\def\mg2{Mg~{\sc ii}}
\def\si2{Si~{\sc ii}}
\def\si4{Si~{\sc iv}}
\def\al2{Al~{\sc ii}}
\def\al3{Al~{\sc iii}}
\def\o1{O~{\sc i}}
\def\n1{N~{\sc i}}
\def\h1{H~{\sc i}}

\def\approxlt{\mathrel{\spose{\lower 3pt\hbox{$\sim$}}
        \raise 2.0pt\hbox{$<$}}}
\def\approxgt{\mathrel{\spose{\lower 3pt\hbox{$\sim$}}
        \raise 2.0pt\hbox{$>$}}}

\documentclass[cup5b]{caps}
\usepackage{graphicx}
\usepackage{amssymb}
\usepackage{ociwsymp4e}  

\HeadText{McWilliam and Rich}  

\def\plotone#1{\centering \leavevmode
\includegraphics[width=.95\columnwidth]{#1}}

\def\plotone#1{\centering \leavevmode
\includegraphics[width=.95\columnwidth]{#1}}

\begin{document}

\pagenumbering{arabic}

%
%


\author[]{ANDREW MCWILLIAM$^{1}$ and R. MICHAEL RICH$^{2}$ 
\\
(1) Carnegie Observatories, Pasadena, California, USA \\
(2) University of California, Los Angeles, Los Angeles, California, USA}

\chapter{Composition of the Galactic Bulge}

\begin{abstract}

We present detailed abundance results for 9 Galactic bulge stars in Baade's
Window, based on HIRES (R=45,000--60,000) spectra taken with the Keck~I telescope.

The alpha elements show non-uniform enhancements relative to the solar neighborhood
trends: Mg and Si are enhanced in all our bulge stars by $\sim$0.5--0.3 dex, showing a slight
decrease with increasing [Fe/H].  Oxygen is enhanced in most bulge stars, similar to the
Galactic halo, but the [O/Fe] ratios suddenly decline beginning at
[Fe/H]=$-$0.5 dex, with a slope consistent with no oxygen production in the bulge for
[Fe/H]$\geq$$-$0.5 dex.  The observed production of magnesium in the absence of oxygen synthesis
appears to be inconsistent with current predictions for supernova nucleosynthesis yields; we
suggest that this may be connected to the Wolf-Rayet phenomenon, which occurs in the pertinent
metallicity range.  The trend of [Ti/Fe] shows an enhancement relative
to the solar neighborhood of $\sim$0.3 dex, but declines somewhat by solar metallicity;
thus, the Ti trend is less extreme than found by McWilliam \& Rich (1994, MR94). 
For metal-poor bulge stars calcium is enhanced by $\sim$0.3 dex, similar to the
Galactic halo, but in our bulge stars with [Fe/H]$\geq$$-$0.5, Ca show a slight deficiency
relative to the solar neighborhood value, consistent with the findings of MR94; this
may indicate a slightly top-heavy IMF as suggested by MR94 and McWilliam (1997).
Aluminum is enhanced in all our Galactic bulge giant stars by $\sim$0.25 dex; we 
argue that this is consistent with nucleosynthesis dominated by type~II
supernova events (SNII), implied by the Mg and Si enhancements.

The trend of the r-process element europium with [Fe/H] is similar to
the [Ca/Fe] trend with metallicity, suggesting that the site of the r-process is 
associated with low-mass SNII, thought to be the dominant producers of calcium.
The odd-Z element Mn shows approximately the same trend with [Fe/H] as seen in the
solar neighborhood; this argues that the solar neighborhood trend is not due
to Mn over production in type~Ia supernova events (SNIa).

\end{abstract}

\section{Introduction}

We have been engaged in a long term program
to study the composition of stars in the Galactic bulge, with 
the aim of constraining the conditions of the bulge's formation
and chemical evolution at a level of detail that is impossible
to obtain with any other method.  Our plan is to obtain high
S/N, high resolution spectroscopy for samples of $>$20 stars in
each of 3 latitudes in the bulge:  Baade's Window ($b$=$-$4$^{\circ}$),
Sgr~I ($b$=$-$2.65$^{\circ}$), and the Plaut field ($b$=$-$8$^{\circ}$).
The Baade's Window sample was obtained using the HIRES spectrograph 
(Vogt et al. 1994) at the W.M. Keck Observatory.  We have completed observations
of Plaut's field using the MIKE spectrograph at the Magellan telescope,
and we hope to complete Sgr~I in 2004.  We also plan to use
the fiber mode of the MIKE spectrograph to obtain
single-order spectra for 200 giants in each of these bulge fields.  
This will enable the homogeneity of the 
chemical properties of the bulge to be investigated, which may reveal
evidence of galactic accretion, or multiple star formation events.
Pertinent questions might include: What are the metallicity and composition
gradients in the Galactic bulge, and how do they constrain bulge formation
scenarios.  Did some fraction of the population form in separate systems that
merged very early on?  What fraction of the population might have originated
in a system similar to the Sagittarius dwarf spheroidal galaxy?  Are there
correlations between kinematics and composition, and do the composition gradients
differ from the iron abundance gradient?

The fossil record of the Local
Group is important because we cannot trace, with certainty, the evolution
of any individual galaxy or galaxy population to the present epoch.
In order to understand the formation of galaxies such as our own we
must also use the constraints available from the ages, kinematics, and
chemistry of stars in its constituent stellar populations. 
This is especially true if present day galaxies 
accumulated through a complicated set of mergers, as suggested in 
the CDM galaxy formation scenarios (e.g.  Kauffmann 1996).
K giants provide useful probes for understanding chemical evolution, because
they are luminous, cover the full range of possible ages, contain lines
from many elements, and except for a few light elements they preserve the 
composition of the gas from which they were formed.
The widely accepted paradigm (Tinsley 1979)
is that yields of massive star Type~II SNe are
enriched in alpha elements (e.g. O, Mg, Si, Ca, Ti), and the 
longer-timescale Type~I SNe are enriched in iron. 
Consequently, the relative importance of alpha elements and iron gives a
constraint on the enrichment timescale for the stellar population.
For example, Elmegreen (1999) models bulge formation as a
maximum intensity star burst that concludes in $10^8$ yr.
The consequences for chemical evolution are modeled by
Matteucci, Romano, \& Molaro (1999), which predicts
abundance trends in the bulge for a wide range of
elements.  The expectation is that the rapid formation
of the bulge should produce an over enhancement of alpha
elements.  However, all elements are of interest because
different families of elements are thought to be made under a variety
of astrophysical circumstances, which can provide information on
the bulge environment during its formation, and may also lead to a greater
understanding of nucleogenesis.
Surveys of large samples of stars in globular clusters
and dwarf spheroidal galaxies are just beginning, and the proposed
research will add the bulge population.

\subsection{A Brief history of Abundance Analysis for Bulge Giants}

The earliest efforts to obtain spectra of bulge stars with linear
detectors were by Whitford \& Rich (1983) and Rich (1988).  These
studies concluded that there is a range of metallicity in the
bulge, and there is evidence for super-metal-rich stars.
Rich's (1988) mean abundance for the bulge was [Fe/H]$\sim$0.2.
McWilliam \& Rich (1994, MR94) obtained marginal high-resolution echelle spectroscopy 
($R$=17,000, S/N$\sim$50)
of 11 bulge K giants, from the sample of Rich (1988), using the CTIO Blanco telescope,
and revised the abundance distribution in the bulge downward.  The results of
MR94 indicated a mean [Fe/H]=$-$0.15, and enhancements of the alpha elements
Mg and Ti, even for stars with the solar iron abundance.  The alpha elements
Si and Ca did not appear to be enhanced over the solar neighborhood trends, and
this was explained by MR94 in the context of the supernova nucleosynthesis predictions
of Woosley \& Weaver (1995, WW95), as due to a top-heavy bulge IMF.
These early studies were made difficult by the lack of accurate basic
optical and infrared photometry for the bulge, and the large, but uncertain, 
reddening (now known to be highly variable).  The high metallicity and cool temperatures of
the bulge giants also posed difficulties due to blending, and because it was necessary to
use lines with somewhat uncertain $gf$ values.  In addition, the extreme stellar crowding
in these low-latitude fields, and the faintness of the bulge stars, made it impossible to
obtain the highest quality
data with 4m-class telescopes; as a result, the use of relatively low resolving power
spectra resulted in considerable blending with CN lines.

The first spectrum
of a bulge giant with Keck (Castro et al. 1996) confirmed the metal-rich end
of the MR94 abundance distribution; however, the spectrum was of low S/N and it
was clear that a serious Keck telescope campaign was urgently needed.
This effort has been underway since August 1998.  Acquisition of
the spectra has proceeded slowly, since we require $S/N>50$ and
$R>45,000$; in fact, most of our spectra have $R=60,000$.  

An analysis of a subset of the new Keck spectra was given in Rich \& McWilliam
(2000, RM00).  The analysis employed model atmosphere parameters based entirely
on the spectra rather than on photometry: effective temperatures were set by
demanding that Fe~I line abundances were independent of excitation potential,
whilst the stellar gravities were chosen so that Fe~I and Fe~II abundances were
equal.  The RM00 abundances at the metal rich end are $\sim$0.15 dex more metal
rich than MR94.  The key results of MR94 were confirmed for the alpha elements: 
[Mg/Fe] and [Ti/Fe] are elevated, while Ca followed the disk abundance
trend.  Oxygen was measured cleanly for the first time, and it was found
that [O/Fe] declines rapidly above [Fe/H]=$-$0.5, a result that we
confirm with the new work presented here.  
One problem with the RM00 study was due to the sensitivity of the spectroscopic
model atmosphere parameters the adopted $gf$ values of the iron lines.  In addition,
the covariance between the spectroscopic temperature and gravity meant that
a slight error in temperature returned an error in gravity, which then
required an additional adjustment in temperature, in the same direction as the
original error.  Thus, it was possible to obtain spectroscopic parameters unexpectedly
far from the true values.

In the present study we have employed two spectroscopic methods
for estimating stellar parameters, and we check these with V$-$K colors
derived from the newly available 2MASS K-band magnitudes together with the
V-band measurements from the OGLE microlensing survey.  These three methods
provide excellent agreement in the stellar parameters, so we are confident
that the present study currently provides the best approach for the detailed 
abundance analysis of bulge giants.  Our results for the iron abundance scale 
now returns more closely to that of MR94, and we give new results on
abundance trends for other elements including some quite puzzling results for oxygen.

\section{Observations and Data Reduction}

Target bulge red giants were selected from the list of Arp (1965), MR94
and Sadler, Rich \& Terndrup (1996).
The spectra for this work were accumulated using the Keck~I telescope and HIRES echelle
spectrograph, over a period from 1998 to 1999 (the complete dataset includes additional
observations and will be discussed by Fulbright, Rich \& McWilliam 2003, in preparation).
For most bulge stars the HIRES C1 decker was chosen, giving a resolving power of R=45,000; 
however, the most metal rich stars were observed using the B2 slit, giving R=60,000.  
Typical S/N ratios were 50:1 per extracted pixel.  The HIRES 2048$^2$ CCD enabled a 
wavelength coverage of $\lambda$$\lambda$ 5400--7800\AA ; although, roughly half of this 
range did not fall on the CCD.  Because of the incomplete wavelength coverage it was 
necessary to select the spectral region carefully, in order to include the most important 
lines.

The spectra were extracted using {\em makee}, a pipeline reduction program
written by T.Barlow.  We preferred not to use the {\em makee}
output normalized by the quartz lamp, because the result gave an unacceptable,
very non-uniform, flattened spectrum.  This failure was likely due to the
difference in illumination afforded by the quartz lamp compared to the stars.
Instead, we chose to use pixel-to-pixel flat fielding of the quartz lamp, but the
echelle blaze function was determined from a fit to the continuum of a
metal-poor star of similar temperature to the program stars.  This resulted in
an improved flattening; however, the results sometimes showed residual artificial 
modulation, so it was necessary to find the standard star spectrum which gave the flattest
result.  We are not sure why this problem occurred, but we suspect that the HIRES field 
rotation and atmospheric dispersion correction optics imposed a modulation of the stellar
flux.  This difficulty could impose systematic effects on our spectra, which would go 
unnoticed for metal-rich stars, which have little continuum.  In Figure~1 we show sample 
reduced spectra.

\begin{figure*}
\centering
\plotone{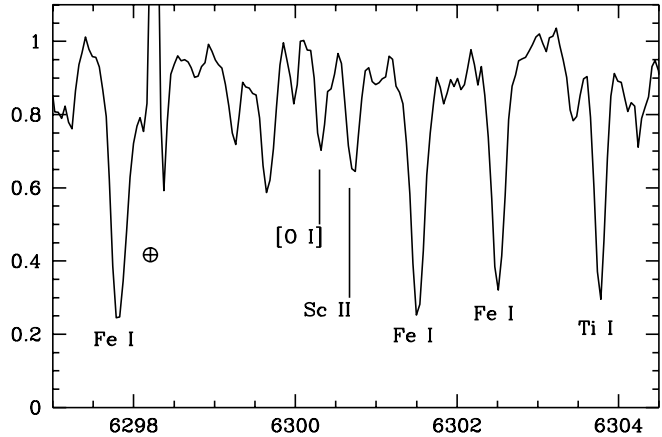}
\vskip0pt
\caption{Keck HIRES spectrum of BW~I-039 (8000 sec exposure).  Notice the clean, unblended,
line of radial velocity-shifted [O~I] 6300\AA .  The high resolution, at R=60,000,
clearly resolves crucial lines, thus permitting abundance measurements for lines
on the linear portion of the curve of growth.  }
\label{o1039}
\end{figure*}

Equivalent widths (EWs) of lines of interest were measured using the semi-automated
program {\em GETJOB}, written by A.McWilliam (McWilliam et al. 1995).  The basic line list
was taken from MR94, with additional lines from Smecker-Hane \& McWilliam (2003).
We note that lines blended in MR94 were often unblended in the present spectra, due to the
significantly higher spectral resolution.

\section{Analysis}

We used the LTE spectrum synthesis program MOOG (Sneden 1973) and the grid
of LTE model atmospheres from Kurucz (1993) to compute abundances
from the measured EWs.  The line list came from MR94, with additional
lines added for the increased wavelength coverage; complete details will
be made available in a later paper.  For Eu and Mn discussed
here we included hyperfine splitting in the abundance calculations; the $gf$
value and $hfs$ parameters for the Eu~II line at 6645\AA\ came from 
Lawler et al. (2001), while the parameters for the Mn~I lines are 
discussed McWilliam et al. (1995), McWilliam \& Smecker-Hane (2004, in
progress), and a later paper.
For the [O I] line at 6300\AA\ we employed the $gf$ values adopted by
Allende Prieto, Lambert \& Asplund (2001), and include their Ni~I blend at
6300.339\AA .  For consistency we re-derived the LTE solar oxygen abundance
by matching the EW of the line in the Kurucz, Furenlid \& Brault (1984) solar atlas,
and found $\epsilon$(O)=8.71; this is only 0.02 dex higher than the solar value of
Allende Prieto et al. (2001) who employed a 3-dimensional time-dependent
hydrodynamical model solar atmosphere in LTE.

\subsection{Model Atmosphere Parameters}

A long-standing difficulty for the study of Galactic bulge stars has been
the paucity of good photometry, useful for the estimation of stellar effective
temperature (T$_{\rm eff}$); even V-band photometry of bulge K giants was
in a poor state until the advent of the OGLE project (e.g. Paczy\'nski et al. 1999).
Although bulge reddening maps have been compiled (e.g. Stanek 1996) there
is a significant variance in reddening over small spatial scales 
(Frogel, Tiede \& Kuchinski 1999).  An additional complication is that the OGLE V$-$I 
colors differ from solar neighborhood stars (e.g. Stutz, Popowski \& Gould 1999;
Kubiak et al. 2002).  Given these difficulties we felt it prudent to
investigate T$_{\rm eff}$ values derived using spectroscopic techniques.

The RM00 work employed Fe~I excitation and Fe~I/Fe~II ionization equilibrium
to determine T$_{\rm eff}$ and log~g respectively; they obtained atmosphere parameters 
and [Fe/H] values significantly higher than found by MR94.  The parameter change from
MR94 was exacerbated by the covariance of the spectroscopic temperatures and gravities:
an increase in T$_{\rm eff}$ led to an increase in log~g, which resulted in an additional
T$_{\rm eff}$ increase.  

For Arcturus Smecker-Hane \& McWilliam (2003, SM03) found a disconcerting 110K difference 
between the physical T$_{\rm eff}$ (based on the known flux, distance and angular 
diameter) and T$_{\rm eff}$ based on Fe~I excitation (T$_{\rm ex}$).  This difference 
was reduced to 40K if a group of high-excitation Fe~I lines was excluded from the 
analysis, and suggests problems with the Fe~I $gf$ values, or blends, at red wavelengths.  
The remaining 40K difference may be due to problems with the upper layers of the model 
atmospheres, where the empirical T-$\tau$ relation for Arcturus (Ayres \& Linsky 1975) 
differs significantly from the Kurucz (1992) model (see also McWilliam et al 1995b).

In addition to excitation temperatures SM03 employed Fe ionization equilibrium to 
determine spectroscopic temperatures; this was made possible by fixing the gravity at 
the physical value (based on the distance and a mass estimated from color-magnitude 
diagram and theoretical evolutionary tracks).  The resultant ionization temperature 
(T$_{\rm ion}$) is likely more robust than the excitation temperature, because the slope
of abundance with excitation is more sensitive to individual erroneous points than the 
average of the Fe~I or Fe~II abundances.  To compute ionization temperatures
the stellar gravities must be determined independently; the required ingredients
for this are the distance, reddening, observed flux and assumed masses.  We
note that $\alpha$ elements are important electron donors in the atmospheres
of red giants, so the [$\alpha$/Fe] ratios should be taken into account in the
calculation of ionization temperatures.  The lesson learned from Arcturus is that 
excitation temperatures are sensitive to systematic $gf$ value errors and blends; this 
is likely more severe for stars more metal-rich than Arcturus.

For the bulge stars analyzed in the present work we computed both excitation and 
ionization temperatures.  We also derived T$_{\rm ex}$ and T$_{\rm ion}$ with abundances
taken relative to Arcturus, on a line by line basis; this ensured that $gf$ value 
uncertainties did not affect the result.  For the bulge K giants systematic errors due 
to problems with the Kurucz (1992) model atmospheres, and line blends, will partially,
or completely, cancel when normalized Arcturus (a K0 giant), and lead to more reliable
excitation temperatures.  For the ionization temperatures we employed physical gravities
based on the observed luminosity and temperature, computed using the dereddened 
OGLE-II V-band magnitudes (Paczy\'nski et al. 1999), a distance of 8.0Kpc 
(e.g. Merrifield 1992, Carney et al. 1995), and an assumed mass of 0.8M$_{\odot}$.

Our adopted stellar atmosphere parameters are listed in Table~1.  Since completing the
current abundance work the 2MASS near-IR catalog has been released for the 
Baade's Window field, which enables temperatures to be computed from the robust V$-$K 
color-temperature calibration of Alonso et al. (1999, 2001).  In Table~1 we include the
V$-$K temperatures, for comparison with the spectroscopic values, computed assuming the
Paczy\'nski et al. (1999) V-band extinction values and the reddening law of Winkler 
(1997).  We note that the Paczy\'nski et al. (1999) (V$-$K) reddening corrections are 
typically 0.18 magnitudes less than the value adopted by MR94 (from Frogel, 
Whitford \& Rich 1984); this would result in effective temperatures $\sim$100K cooler 
from the Paczy\'nski et al. photometry.  The mean difference between spectroscopic and
(V$-$K) temperatures, for the 7 stars in Table~1, is 49K, with a standard deviation about
the mean difference of 50K; thus, the two temperature scales are in good agreement.

\begin{table}
\caption{Galactic Bulge Star Atmosphere Parameters}
\begin{tabular}{rcccccc}
\hline
\hline
    Star   & [Fe/H]$_{\rm lines}$ &
       T$_{\rm eff}$ & logg & [Fe/H]$_{\rm model}$ &  $\xi$ (km/s) & T$_{\rm eff}$(V$-$K) \\

\hline

 I--194 &  $-$0.09  & 4210 &  1.9 &  $-$0.1  &  1.04 &  4170 \\
 I--202 &  $+$0.02  & 4200 &  1.6 &  $+$0.0  &  1.50 &  4229 \\
 I--322 &  $-$0.12  & 4190 &  1.6 &  $-$0.1  &  1.50 &  4176 \\
 II--166 &  $-$1.60  & 5010 &  1.4 &  $-$1.6  &  1.59 &   ... \\
III--152 &  $-$0.32  & 4190 &  1.8 &  $-$0.4  &  1.19 &  4171 \\
 IV--003 &  $-$1.28  & 4500 &  1.5 &  $-$1.3  &  1.43 &  4431 \\
 IV--072 &  $+$0.28  & 4250 &  1.9 &  $+$0.3  &  1.42 &   ... \\
 IV--167 &  $+$0.35  & 4360 &  2.0 &  $+$0.4  &  1.54 &  4305 \\
 IV--203 &  $-$1.22  & 3920 &  0.5 &  $-$1.2  &  1.75 &  3833 \\
 IV--329 &  $-$0.84  & 4300 &  1.5 &  $-$0.9  &  1.35 &  4165 \\
\hline
\end{tabular}
\end{table}

\section{Abundance Results and Discussion}

The [Fe/H] results from the current work are only 0.03 dex, on average, higher 
than found by MR94, with a dispersion about the mean difference of 0.21 dex.
If the current work and MR94 contribute equally to the dispersion each has a
characteristic uncertainty of 0.15 dex; however, the errors are probably
larger for the MR94 analysis.  The RM00 [Fe/H] values were 0.14 dex higher
than MR94; thus our sample is 0.11 dex lower than the RM00 metallicity scale.
The difference between the current work and RM00 is due to our improved
method for determining spectroscopic model atmosphere parameters, which is
less prone to systematic errors than the classical method employed by RM00.

In Figure 2 we show the abundance ratios for 3 $\alpha$-elements: O, Mg and Si.  
The [Mg/Fe] and [Si/Fe] ratios appear enhanced at $\sim$0.3 dex at all bulge [Fe/H],
similar to the values seen in the halo, and well above the Edvardsson et al. 
(1993) trend of [$\alpha$/Fe] for local disk stars (solid line in Figure~2).
These enhancements in Mg and Si suggest that SNII played a
dominant role in the synthesis of the elements at all [Fe/H] in the bulge, and that
the bulge formed rapidly (e.g. Matteucci et al. 1999).  The result confirms the 
Mg enhancement found by MR94, but differs from their claim of an apparently normal [Si/Fe]
trend with [Fe/H].  Given the superiority of the current spectra, and analysis methods,
we put greatest weight on the current Si results.

\begin{figure*}
\centering
\plotone{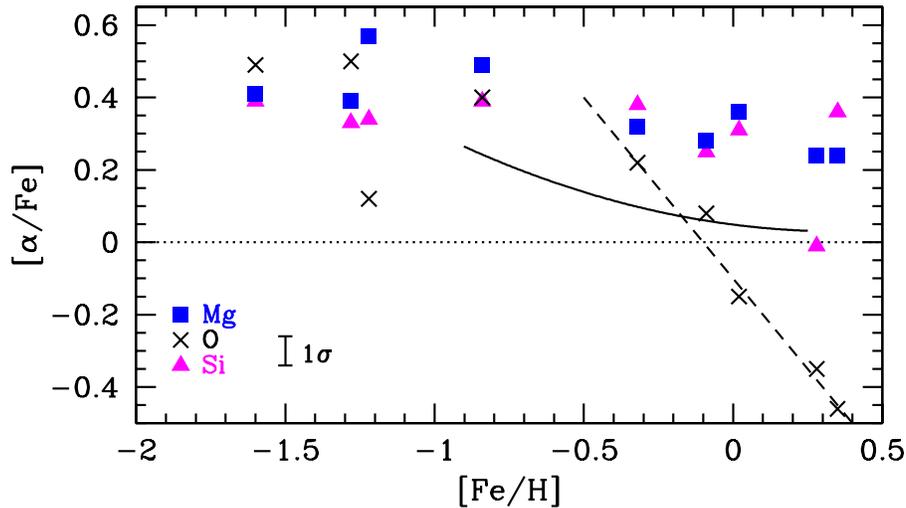}
\vskip0pt
\caption{The trend of alpha elements [Mg/Fe] (filled squares), [Si/Fe] (filled triangles),
and [O/Fe] (crosses), versus [Fe/H], in the Galactic bulge.  The solid line represents
the [$\alpha$/Fe] trend of Edvardsson et al. (1993).  The dashed line indicates the
locus of constant [O/H] above [Fe/H]=$-$0.5 dex.}
\label{mgsio2}
\end{figure*}

Figure~2 show that the oxygen abundances are enhanced by $\sim$0.4 dex for bulge stars 
with [Fe/H]$\leq$$-$0.8, similar to the values seen in the Galactic halo, but slightly 
higher than the non-oxygen $\alpha$-element enhancements most metal-poor disk stars of 
Edvardsson et al. (1993).  The figure also shows a strong decline in [O/Fe] with 
increasing [Fe/H] for bulge stars with [Fe/H]$\ge$$-$0.5; the slope is consistent with 
no production of oxygen above [Fe/H]=$-$0.5 (indicated by the dashed line).

It is significant that the sharp decline in [O/Fe] is not seen in [Mg/Fe], because both 
Mg and O are thought to be produced mostly in massive SNII events.  Slightly enhanced 
[Mg/O] ratios are seen in some thin disk stars observed by Bensby et al. (2003), but 
the difference is smaller than found here.  The SNII yields of WW95 
predict an increased production of oxygen relative to magnesium with increasing 
progenitor mass; but in no case is Mg produced without O.  One might appeal to 
metallicity-dependent yields and the Wolf-Rayet effect to understand the bulge O/Mg
production; certainly, the Wolf-Rayet phenomenon affects massive SNII progenitors near
[Fe/H]=$-$0.5.  Qualitatively, one might expect greatly reduced hydrostatic oxygen 
production if the masses of the SNII progenitors is significantly reduced by Wolf-Rayet
winds.  However, predicted nucleosynthesis yields of exploding Wolf-Rayet stars by
Woosley, Langer \& Weaver (1995) indicate solar O/Mg ratios.  Thus, we are unable to
explain the observed halt in oxygen production at [Fe/H]=$-$0.5, and put it forward as
a challenge for nucleosynthesis theorists.

If the [Fe/H] distribution of Sadler, Rich \& Terndrup (1996) is taken at face value, 
then 80\% of the Baade's Window stars are more metal-rich than [Fe/H]=$-$0.5, the 
point where oxygen production in the bulge stopped.

In Figure~3 we show plots of [Ti/Fe], [Ca/Fe] and [O/Fe] with [Fe/H].  The [Ti/Fe] 
ratio is enhanced in most bulge stars by $\sim$0.3 dex, similar to the Galactic halo;
at metallicities greater than solar the general level of the Ti enhancement is uncertain,
but appears reduced.  In a very careful study Reddy et al. (2003) indicate that
[Ti/Fe]$\sim$0.0 for the thin Galactic disk, roughly 0.1 dex lower than the
Edvardsson et al. (1993) trend; thus, if we compare our [Ti/Fe] ratios with the 
Reddy et al. results the bulge appears enhanced at all [Fe/H].
This conclusion is similar to, but weaker than, the the Ti enhancements found by MR94,
and suggests that SNII played a more important role in bulge nucleosynthesis than for
the disk; however, since the nucleosynthesis of Ti is not well understood the 
conclusions which can be drawn are limited.

  \begin{figure*}
    \centering
    \plotone{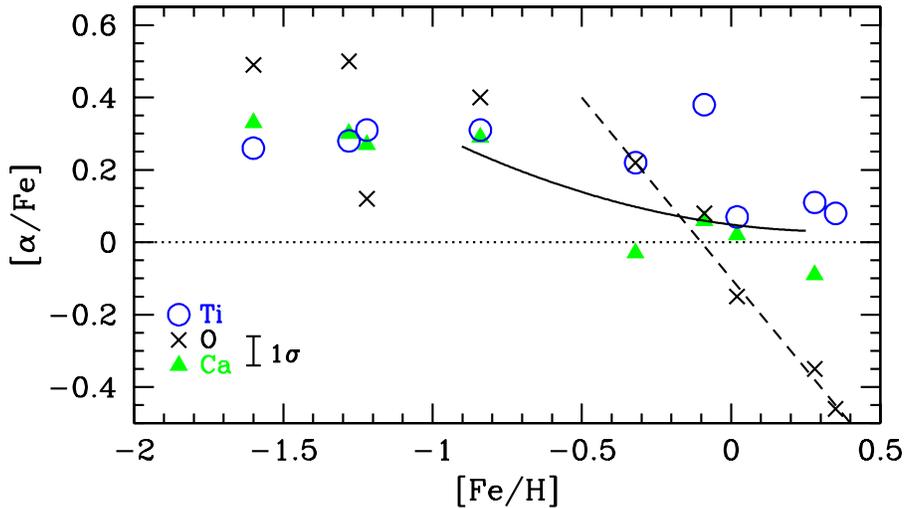}
\caption{The trend of alpha elements [Ca/Fe] (filled triangles), [Ti/Fe] (open circles),
and [O/Fe] (crosses), versus [Fe/H], in the Galactic bulge.  Lines are same as in Figure~2.
The [Ca/Fe] trend is close to the solar value for [Fe/H]$\geq$$-$0.5 dex.}
\label{catiofe}
\end{figure*}

The [Ca/Fe] trend found here is similar to that found by MR94: for [Fe/H]$\leq$$-$0.8
[Ca/Fe] is halo-like, being enhanced by $\sim$0.3 dex; but the ratio is
significantly reduced for more metal-rich stars.  In our sample the average [Ca/Fe] 
ratio for stars with [Fe/H]$\geq$$-$0.5 dex is sub-solar, in contrast to the observed 
enhancement in [Mg/Fe] and [Si/Fe].   These observations are remarkably similar to the 
high [Mg/Fe] and low [Ca/Fe] ratios found for giant elliptical galaxies, with 
abundances measured from integrated-light spectra (e.g. Worthey 1998; Saglia et al. 
2002), which supports our finding and suggest an evolutionary similarity between bulges
and giant ellipticals.

Our bulge [Eu/Fe] values are compared to the disk and halo results from
various studies in Figure~4.  In the solar system composition the abundance
of Eu is consistent with a 97\% r-process fraction.  Europium is often reported as
being an r-process element, by virtue of its large neutron capture cross
section.  We note that large [Eu/Fe] enhancements can occur via the s-process,
but in this case the enhancements of Ba, La and other strongly s-process
elements will be even greater; RM00 showed that the [Ba/Eu] ratio for bulge
stars is consistent with the halo mixture, dominated by the r-process.  Like the halo
the metal-poor bulge stars show a $\sim$0.4 dex enhancement in [Eu/Fe], and is consistent
with the notion that Eu, like $\alpha$ elements, are produced by SNII events.  For 
the more metal-rich bulge stars, [Fe/H]$\geq$$-$0.5, [Eu/Fe] appears roughly constant
near the solar value.  The most metal-rich bulge star in this sample, 
IV~167, may be enhanced by $\sim$0.2 dex; but this could possibly be due
to a contaminating blend, most likely from CN, with the Eu~II line at 6645\AA .
It is notable that the [Eu/Fe] trend does not seem to continue declining, as
seen in the solar neighborhood disk stars.  The [Eu/Fe] ratio appears flat
with [Fe/H] in the metal-rich bulge stars; in this regard Eu behaves more like Ca
than the other $\alpha$-elements, which suggests a connection in the formation
of these two elements.  Since theoretical nucleosynthesis yields (e.g. WW95)
indicate that Ca is produced mainly by low-mass (15--25 M$_{\odot}$)
SNII progenitors the similarity between the [Eu/Fe] and [Ca/Fe] trends
suggests that it is in this mass range that the r-process is most important.
Current theoretical predictions have not yet identified the source of the
r-process, so our observations of the bulge composition is of special interest.
In this regard it is particularly important to measure the abundances of
r-process and $\alpha$-elements in different galactic environments, as the
correlations may yield information on the r-process site.

  \begin{figure*}
    \centering
    \plotone{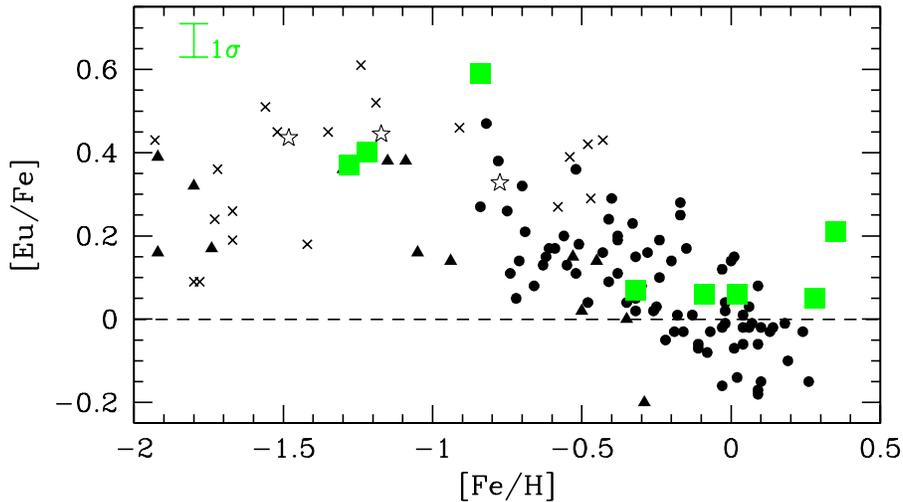}
\caption{The trend of the r-process neutron-capture element europium with
[Fe/H] (large filled squares) for stars in the Galactic bulge.
Results for the disk and halo are shown for comparison: filled circles from
Woolf, Tomkin \& Lambert (1995), filled triangles from Gratton \& Sneden (1994), and
crosses from Shetrone (1996);  open stars indicate the mean values for
globular clusters M71, M13, M5, and M92 from Shetrone (1996).  }
\label{eufehfs}
\end{figure*}

Our results for [Al/Fe] are shown in Figure~5, and compared to the solar
neighborhood results of Chen et al. (2000) and the Sagittarius dwarf
spheroidal galaxy from Smecker-Hane \& McWilliam (2003).  It is clear that 
the value of [Al/Fe] is sensitive to the galactic environment, and appears to 
be enhanced in more rapidly evolving systems; this is consistent with the
production of Al in SNII events, as expected from nucleosynthesis theory.
Our mean [Al/Fe] ratio, at $+$0.25 dex is identical to the results of Reddy et
al. (2003) for thick disk stars in the Galaxy.  This
supports our findings for the bulge, which like the thick disk, is thought
to have formed on a rapid timescale.  
A plot of [Al/Mg] versus [Fe/H] does show a slope, qualitatively in agreement
with expectations of metallicity-dependent yields from SNII (e.g. Arnett 1971).
Near [Fe/H]=$-$1 the bulge [Al/Mg] ratio is very similar to the predictions of
WW95; however, the observed bulge slope of [Al/Mg] with [Fe/H], at $\sim$0.12 dex/dex,
is much shallower than the value of $\sim$0.35 dex/dex indicated in the calculations of
WW95.  At solar [Fe/H] the
WW95 predictions for [Al/Mg] are $\sim$0.25 dex higher than our bulge observations.
Part of the discrepancy is likely due to the unusual composition of the bulge, which
complicates the comparison between WW95 and our observations.

  \begin{figure*}
    \centering
    \plotone{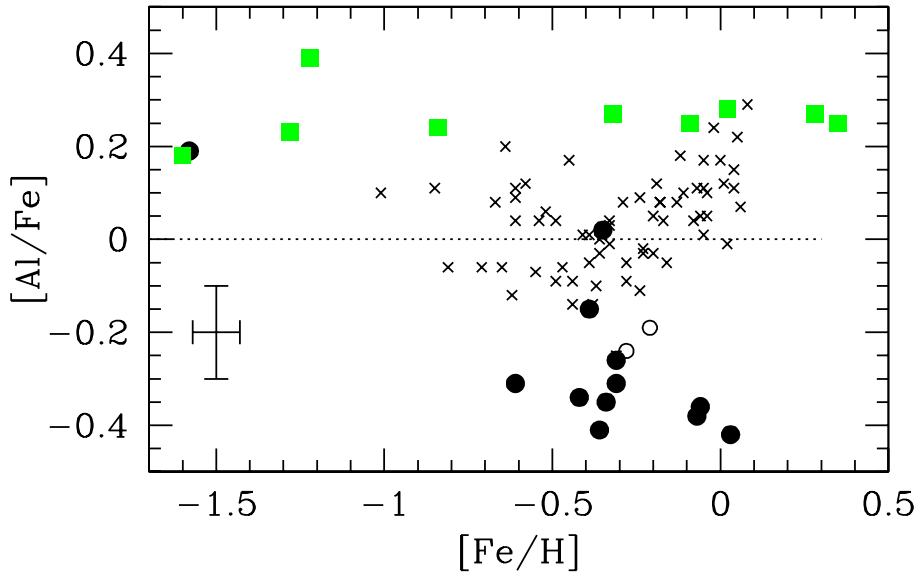}
\caption{
A comparison of [Al/Fe] for stars in the Galactic bulge (filled squares),
the solar neighborhood (crosses from Chen et al. 2000), and the Sagittarius dwarf
spheroidal galaxy (filled circles from Smecker-Hane \& McWilliam 2003; open
circles from Bonifacio et al. 2000); clearly, [Al/Fe] depends upon galactic 
environment.
}
\label{alfe}
\end{figure*}

Our enhanced [Al/Fe] ratios are unlikely to be related to the enhancements
seen in some globular cluster red giant stars (e.g. see Sneden, Ivans \& Fulbright
2003, this conference).  The globular cluster enhancements, due to proton
burning, are characterized by large star-to-star differences, unlike the
roughly constant [Al/Fe] values seen in the bulge; also unlike the bulge, the 
globular cluster 
Al enhancements are accompanied by depletions of oxygen.  Furthermore, the
general consensus is that globular cluster Al enhancements are associated with 
high stellar density, because no field stars have been found with 
abundance patterns associated with Al-enhancement in globular clusters; although
the bulge appears crowded, its density is far below that of globular clusters.

In Figure~6 our abundance trend of [Mn/Fe] with [Fe/H] is shown, and compared to the 
solar neighborhood results, with data from numerous studies (see McWilliam, Rich \&
Smecker-Hane 2003); it is clear that Mn behaves similarly in the bulge and solar
neighborhood.  The observed increase in [Mn/Fe] above [Fe/H]$\sim$$-$1 appears entirely 
consistent with the nucleosynthesis predictions for SNII by WW95.

  \begin{figure*}
    \centering
    \plotone{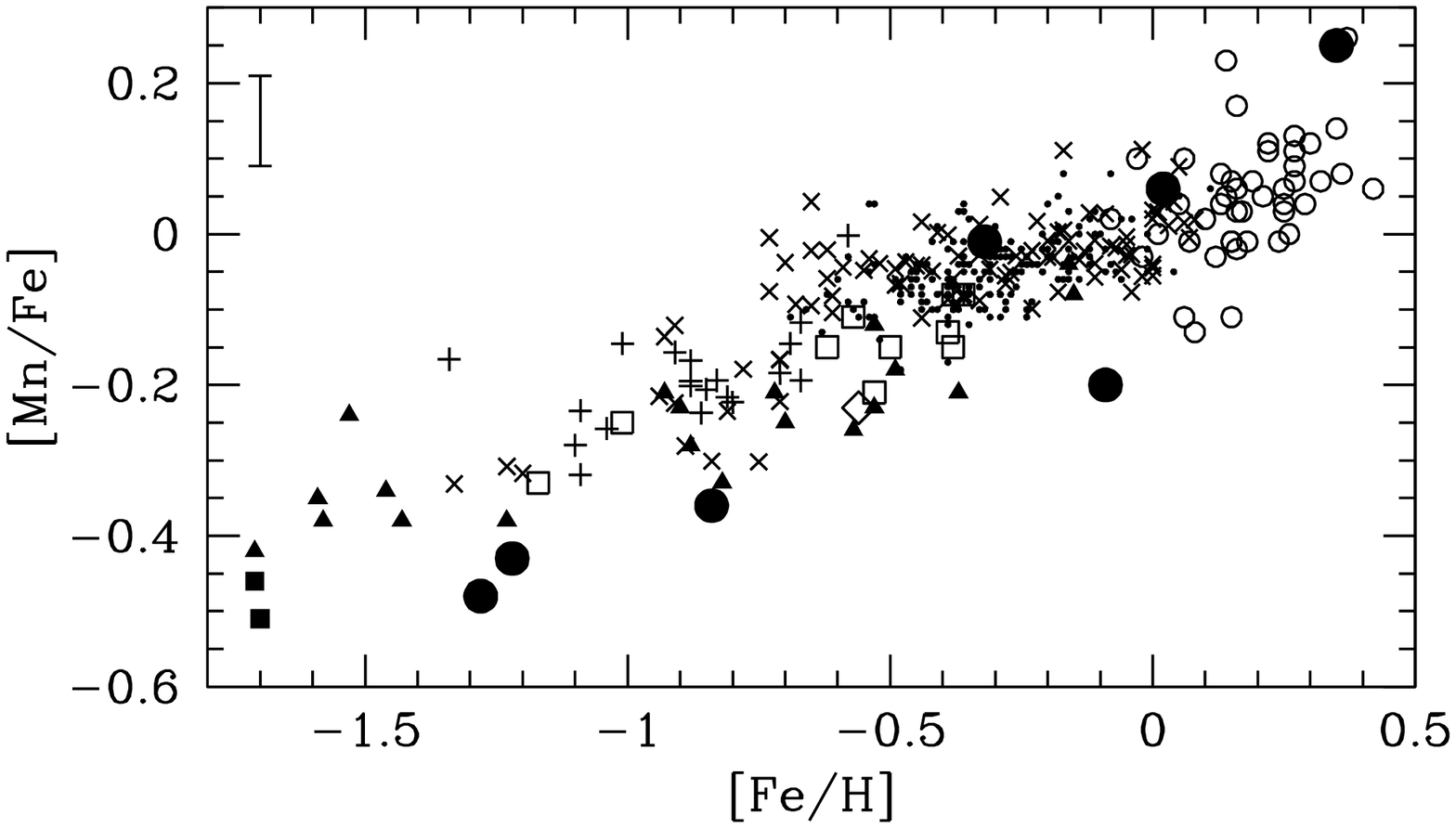}
\caption{ The trend of [Mn/Fe] with [Fe/H] in the galactic bulge stars
(large filled circles) compared to the solar neighborhood (see McWilliam et al. 2003
for the list of sources).  The bulge points approximately follow the solar-neighborhood
trend; large deficiencies might be expected if Mn is over-produced in type~Ia supernovae,
as suggested by Gratton (1989).}
\label{mnfe}
\end{figure*}

Our bulge Mn abundances are not consistent with the notion that the Mn trend
in the solar neighborhood is due to Mn over-production by 
SNIa, as proposed by Gratton (1989).  Gratton noted that the form of
[Mn/Fe] relation with [Fe/H] appears as the inverse of the trend of $\alpha$-elements
with metallicity, and proposed that the increase in [Mn/Fe] was due to the onset of
SNIa, which produce excess Mn.  If the Gratton scenario were correct, then
systems like the bulge, characterized by rapid formation, with little time for 
significant nucleosynthesis by SNIa, would show deficient [Mn/Fe] ratios compared 
to the solar neighborhood trend.  Our conclusion against the Gratton scenario is 
supported by deficiencies in [Mn/Fe] seen in the Sagittarius dwarf spheroidal
galaxy (McWilliam et al. 2003).

It is of value to compare our results with the composition found for stars in the
bulge globular cluster NGC~6528, by Carretta et al. (2001, C01):  Both studies indicate general
enhancements of $\alpha$-elements, but C01 find Ca enhanced by $\sim$0.2 dex, whereas we find
no Ca enhancement above [Fe/H]$\sim$$-$0.5.  To investigate this difference we computed
abundances from the published C01 equivalent widths and atmosphere parameters; our 
calculations show
good agreement for Fe lines, but $\sim$0.2 dex deficiency in [Ca/Fe] relative to
the sun; thus, either C01 included a sign error, or their analysis led to the
different result.  We note that the Ca lines in C01 are strong enough to be quite
saturated, and thus the abundances are quite dependent on the treatment of the
microturbulent velocity.  The low Ca/Fe result found in this paper include stars
more metal deficient than C01, with more lines on the linear portion of the curve
of growth, and less likely to suffer from problems with microturbulent velocity.
We note that C01 found enhanced [Na/Fe], similar to the [Al/Fe] result reported
here; it is entirely expected that Na should follow Al when Al is produced by SNII,
as we propose.  Our bulge results do not confirm the low [Mn/Fe] found by C01 for NGC~6528,
despite the inclusion of Mn $hfs$ treatment; if this difference is confirmed
then NGC~6528 evolved in a very different manner than the Galactic bulge.
Low [Mn/Fe] near solar [Fe/H] for NGC~6528 could occur if the cluster were
formed from the ejecta of many metal-poor SNII, resulting in low, halo-like,
[Mn/Fe] at higher than halo [Fe/H].

\section{Conclusions}

We find that the alpha elements Mg, Si and Ti are enhanced in the Galactic
bulge, consistent with nucleosynthesis by type~II supernovae, even at solar
metallicity.  Oxygen is enhanced in the metal-poor bulge, but above
[Fe/H]$\sim$$-$0.5 the [O/Fe] ratio declines steeply, with a slope consistent
with no oxygen production.  The decline in [O/Fe] without a similar decline
in [Mg/Fe] is a challenge for supernova nucleosynthesis theory.  We naively
suggest that the onset of the Wolf-Rayet phenomenon may be connected to
the cessation of oxygen production in the bulge.

The alpha element Ca and the (mostly) r-process element Eu are enhanced in the
metal-poor bulge, but lie near the solar value at metallicities above [Fe/H]=$-$0.5.
The low calcium abundances are similar to the composition of giant elliptical
galaxies and might be understood as due to a paucity of low-mass SNII
progenitors, due to a top-heavy IMF.  The similarity of the [Eu/Fe] and [Ca/Fe]
trends suggest a connection in the formation of these two elements, indicating that
the r-process occurs mainly in low-mass SNII.  The correlation of r-process and alpha 
element abundances presents an interesting way to constrain the site of the r-process,
which should be exploited further.

The [Al/Fe] ratio is enhanced in the Galactic bulge, similar to results for
the thick disk, and is consistent with Al over-production in rapidly evolving
systems.  Although the [Al/Fe] ratio remains constant, near $+$0.25 dex, over a 2 dex
range in [Fe/H], the [Al/Mg] ratio shows a small slope with [Fe/H] in approximate accord
with supernova nucleosynthesis predictions of metallicity-dependent Al/Mg yields;
however, the slope is much less than predicted by WW95.  
Thus, it would be interesting to see whether the observed [Al/Fe] enhancements and [Al/Mg] trend
persist for metal-poor bulge stars well below 
[Fe/H]=$-$1.6, as a test for supernova nucleosynthesis models.

The iron-peak element manganese shows nearly the same trend with [Fe/H] as the solar
neighborhood disk, in accord with the nucleosynthesis predictions of WW95.
This trend would not be expected from scenario of Mn-overproduction in SNIa,
which Gratton (1989) proposed to explain the observed solar neighborhood
[Mn/Fe] trend with [Fe/H].

\end{document}